\documentclass[aps,twocolumn,superscriptaddress,floatfix,letter]{revtex4}

\usepackage{graphicx}
\usepackage{amsmath}
\usepackage{hyperref}
\usepackage{url}
\usepackage{float}

\usepackage{color}

\newcommand{\ket}[1]{\left\vert #1\,\right\rangle}
\newcommand{\bra}[1]{\left\langle #1\,\right\vert}

\begin{document}

\setlength\abovedisplayskip{5pt}
\setlength\belowdisplayskip{5pt}
\title{Exponentially fast dynamics of chaotic many-body systems}
\author{Fausto Borgonovi}
\affiliation{Dipartimento di Matematica e
  Fisica and Interdisciplinary Laboratories for Advanced Materials Physics,
  Universit\`a Cattolica, via Musei 41, 25121 Brescia, Italy}
\affiliation{Istituto Nazionale di Fisica Nucleare,  Sezione di Pavia,
  via Bassi 6, I-27100,  Pavia, Italy}
\author{Felix M. Izrailev}
\affiliation{Instituto de F\'{i}sica, Benem\'{e}rita Universidad Aut\'{o}noma
  de Puebla, Apartado Postal J-48, Puebla 72570, Mexico}
\affiliation{Dept. of Physics and Astronomy, Michigan State University, E. Lansing, Michigan 48824-1321, USA}
\author{Lea F. Santos}
\affiliation{Department of Physics, Yeshiva University, New York,
New York 10016, USA}

\date{\today}

\begin{abstract}
We demonstrate analytically and numerically that in isolated quantum systems of many interacting particles, the number of many-body states participating in the evolution  after a quench  increases exponentially in time, provided the eigenstates are delocalized in the energy shell. The rate of the exponential growth is defined by the width $\Gamma$ of the local density of states (LDOS) and is associated with the Kolmogorov-Sinai entropy for systems with a well defined classical limit. In a  finite system, the exponential growth eventually saturates due to the finite volume of the energy shell. We estimate the time scale for the saturation and show that it is much larger than $\hbar/\Gamma$.  Numerical data  obtained for a two-body random interaction model of bosons and for a dynamical model  of interacting spin-1/2 particles  show excellent agreement with the analytical predictions. 
\end{abstract}

\maketitle

{\em Introduction.--} 
After decades of intensive studies, the term ``quantum chaos''~\cite{CCFI,bil,valz,BGS,CIS,qc,sred,rep1,rep5,our} became widely disseminated and accepted in modern physics. Originally, it referred to   quantum systems whose classical counterparts are  chaotic. Paradigmatic examples are the kicked rotor model (KRM)~\cite{CCFI,CIS}  and billiard models~\cite{bil,valz,BGS}, both of which reveal  quantum signatures of classical chaos~\cite{chirikov,bunimovich}.  It was conjectured  and numerically proved~\cite{valz,BGS} that quantum chaos might be  quantified by specific properties of the fluctuations of energy spectra. In particular, it was  found that in chaotic systems, the distribution of spacings between neighboring energy levels follows closely the Wigner surmise~\cite{footMehta}, in contrast with the Poisson dependence that emerges in integrable systems. 

Throughout the development of one-body quantum chaos, dynamics has played a crucial role. Numerical studies of the KRM~\cite{CCFI,CIS} discovered the unexpected existence of two time scales associated with the quantum-classical correspondence.  It was confirmed that a complete correspondence between the quantum and  classical behavior   occurs only on a tiny time scale according to the Ehrenfest theorem. It was analytically shown in~\cite{bz} that this time scale  is given by $t_E \simeq \lambda^{-1} \ln(I/ \hbar)$, where $I$ represents a characteristic action and $\lambda$ is the classical Lyapunov exponent. However, numerical data reported and discussed in~\cite{CCFI,CIS}  revealed the existence of a much larger time scale on which the behavior of classical and quantum  global observables are equivalent. This time scale was found to be  
$t_D \propto D/\hbar^2$, where $D$ is the classical diffusion coefficient in the momentum space. After such time and in contrast with the classical case, quantum diffusion ceases. This phenomenon, called dynamical localization, was explained by the localization of the eigenstates in momentum space according to the relation $\ell \propto D$, where $\ell$ is the localization length~\cite{CIS,dima}. It was later argued that the dynamical localization   found in the KRM  can be also thought in terms of Anderson localization in pseudo-random potentials~\cite{FGP}. 

Contrary to  one-body quantum chaos,  in quantum many-body systems (MBS),  level statistics is less informative than the structure of the eigenstates in a physically chosen basis~\cite{our,int}. It is now understood, for example, that the relaxation of a quantum MBS to its thermal state requires the presence of chaotic eigenstates~\cite{rep1,rep5,our,lev}. The relaxation of a quantum MBS in the thermodynamic limit has been discussed before~\cite{bog}, but the time scale on which it occurs in finite systems is still an open question. To address this problem, we analyze the relaxation of observables of quantum MBS in  the many-body space.

We consider the quench dynamics described by a Hamiltonian $H=H_0+V$ in the region of parameters where the eigenstates are fully delocalized in the energy shell defined by the inter-particle interaction $V$~\cite{vict,chavda,lea,int,bmi17}.  Specifically, we  prepare the system in a single (unperturbed) eigenstate of $H_0$ and study how the state spreads in the unperturbed many-body basis due to $V$. With the use of a semi-analytical approach, we show that the effective number of unperturbed states participating in the dynamics of quantum MBS increases exponentially in time.  

We find that the exponential growth saturates at a time much larger than the characteristic time $\hbar/\Gamma$ of the initial state decay, where $\Gamma$ is the width of the local density of states (LDOS). [The LDOS describes the energy distribution of the initial state. It is obtained by projecting the initial state on the energy eigenbasis.]  We discuss the physical meaning of this novel  time scale in connection with the quantum-classical correspondence for chaotic MBS and  with the problem of thermalization in isolated quantum MBS. Our analytical estimates are fully confirmed by numerical data obtained via exact diagonalization for two different systems: a  model of randomly interacting   bosons and  a  one-dimensional  (1D) system of spins $1/2$ with deterministic couplings.

{\em Models.--} 
 In   both models,  $H_0$ describes the non-interacting particles (or quasi-particles),  while their interaction is contained in  $V$.
The first model represents  $N$ identical bosons occupying $M$ single-particle levels specified by random energies $\epsilon_s$ with a mean spacing $\langle \epsilon_s- \epsilon_{s-1} \rangle = 1 $ setting the energy scale. The   Hamiltonian reads
\begin{equation}
  H= \sum \epsilon_s \, a^\dag_s a_s  +
 \sum V_{s_1 s_2 s_3 s_4} \, a^\dag_{s_1} a^\dag_{s_2} a_{s_3} a_{s_4},
\label{ham}
\end{equation}
where $a_s^{\dagger}$ ($a_s$) is the creation (annihilation) operator on level $s$, and the two-body matrix elements $ V_{s_1 s_2 s_3 s_4} $ are random Gaussian entries with  zero mean and variance $v^2$. The interaction conserves the number of bosons and connects many-body states that differ by changing at most two particles. This two-body interaction (TBRI) random  model was introduced in~\cite{TBRI,brody} to model nuclear systems. It has been extensively studied for fermions~\cite{vict,alt} and bosons~\cite{bosons}. It has also been used to describe non-random systems, such as the Lieb-Liniger model~\cite{LL} largely investigated experimentally~\cite{rmpLL}. The unperturbed many-body eigenstates $\ket{k} $ of $H_0= \sum_k {\cal E}_k \ket{k} \bra{k} $  are obtained by all possible combinations of   $N$ bosons in $M$ single-particle energy levels according to standard statistical rules. This generates $ {\cal D} = \frac{(N+M-1)!}{N!(M-1)!}$ unperturbed many-body states. The eigenstates $\ket{\alpha}$ of the   Hamiltonian $ H = \sum_\alpha E^\alpha \ket{\alpha} \bra{\alpha} $ are represented in terms of  the  states $|k\rangle$ as  $\ket{\alpha} = \sum_k C_k^{\alpha} \ket{k}$.  

The other model studied has no random terms. It describes a dynamical system of interacting spins-1/2 on a 1D lattice of length $L$. Spin systems are intensively studied in experiments with nuclear magnetic resonance platforms~\cite{nmr} and ion traps~\cite{traps}, as well as similar systems with cold atoms~\cite{cold}. The Hamiltonians $H_0$ and $V$   are given by 
\begin{eqnarray}
H_0 &=&  \frac{J}{4}\sum_{s} \left( \sigma_s^x \sigma_{s+1}^x + \sigma_s^y \sigma_{s+1}^y +\Delta \sigma_s^z \sigma_{s+1}^z \right) , \\
V &=& \lambda \frac{J}{4} \sum_{s} \left( \sigma_s^x \sigma_{s+2}^x + \sigma_s^y \sigma_{s+2}^y +\Delta \sigma_s^z \sigma_{s+2}^z \right),
 \label{ham1} 
\end{eqnarray}
where $\sigma^{x,y,z}_s$ are the Pauli matrices on site $s$. The coupling constant $J=1$ sets the energy scale, $\Delta$ is the anisotropy parameter, and $\lambda$ is the ratio between nearest-neighbor and next-nearest-neighbor couplings~\cite{leajmp}.  The  Hamiltonian conserves the total spin in the $z$-direction, ${\cal S}^z = \sum_{s=1}^L \sigma_s^z/2$, which is here fixed  to ${\cal S}^z=-1$, that is $L$ is even and the number of up-spins (excitations) is given by $N=L/2-1$. The dimension of the Hamiltonian matrix  is  $\frac{L!}{N!(L-N)!}$
When  $V=0$, the model is integrable, while as $\lambda$ increases, it becomes chaotic~\cite{int}.

{\em Basic relations.--} We analyze the wave packet dynamics in the unperturbed basis $\ket{k}$ after switching on the interaction $V$. The system is initially prepared in a particular unperturbed state  $\ket{k_0} $,
\begin{equation}
\ket{\psi(0)}  = \sum_\alpha C_{k_0}^\alpha \ket{\alpha}.
\end{equation}

 The probability to find the evolved state in any basis state  $\ket{k}$ at the time $t$ is 
\begin{equation}
P_k(t) = |\langle k |\psi (t) \rangle |^2 = \sum_{\alpha,\beta}  C_{k_0}^{\alpha \ast} C_{k}^{\alpha } C_{k_0}^\beta C_{k}^{\beta \ast} e^{-i(E^\beta-E^\alpha )t} ,
\end{equation}
which can be written as the sum of a diagonal part, $P_{k}^d=\sum_{\alpha}  |C_{k_0}^{\alpha}|^2 |C_{k}^\alpha|^2$, and an oscillating time-dependent part, $P_{k}^f (t)=\sum_{\alpha\ne \beta}  C_{k_0}^{\alpha \ast} C_{k}^{\alpha } C_{k_0}^\beta C_{k}^{\beta \ast} e^{-i(E^\beta-E^\alpha )t}$.
After a long time and assuming a non-degenerate spectrum, $P_k^f$ cancels out on average and only the diagonal part $P_{k}^d$ survives.

With $P_k(t)$, we construct the quantity of our main interest, the number of principal components,
\begin{equation}
\label{s-ipr}
 N_{pc}(t) = \left\{ \sum \limits_{k} \left[P_{k}^d + P_{k}^f (t)\right]^2  \right\}^{-1},
\end{equation}
also known as participation ratio~\cite{referee}. It measures the effective number of unperturbed states $\ket{k}$ that composes the evolved wave packet. For weak interaction,  $N_{pc}(t)$ oscillates in time. Our focus is, however,  on 
 strong values of $V$, where  $N_{pc}(t)$  increases smoothly and eventually saturates to its infinite time average given by
\begin{equation}
\label{t-ipr}
\overline{N_{pc}^{\infty}} =  \left[ 2 \sum_k (P_{k}^d)^2  -  \sum_{\alpha}  | C_{k_0}^\alpha |^4 \sum_k |C_{k}^\alpha|^4 \right]^{-1}.
\end{equation}
This determines the total number of unperturbed many-body states inside the energy shell.

{\em Dynamics in many-body space.--}  A distinctive property of the dynamics of a quantum MBS  is that it cannot  be described as either ballistic or diffusive in the many-body space.  A pictorial demonstration of how the initial state  spreads  in the many-body space  is given in the Supplemental Material (SM) \cite{sm}. Specifically, on a small time scale, only the basis states directly coupled to the initial state are excited. Their number is much smaller than the total number of basis states, due to the sparse structure of the Hamiltonian matrix.  As  time passes  more basis states are populated inside the shell, until its ergodic filling. This takes place provided the perturbation $V$ is sufficiently strong, so that the eigenstates of $H$ are delocalized in the energy shell. 

To describe the time dependence of $N_{pc}(t)$, we develop a cascade model to monitor the flow of probability to find the system in specific unperturbed states at different time steps. This is done by dividing the dynamical process in different time intervals associated with different sets of basis states (classes). At $t=0$, only the ${\cal M}_0$ class is not empty:  it has one element, which is the initial state $\ket{k_0}$. In the next time step, all states having a non-zero coupling with the initial basis state are populated. That is, the first class ${\cal M}_1$  contains the basis states $|k\rangle$ for which $\langle k_0 | V | k\rangle \ne 0$. The second class ${\cal M}_2$ consists of those states which have non-zero matrix elements with all states from the first class. In the same manner, one can define all classes in the many-body space. 

For an infinite number of particles, there is an infinite hierarchy of equations describing the flow of probability from one class to the next one. However, for the values of $N$ and $M$ accessible to our computers, the number of states in the second class practically coincides with ${\cal D}$, so only two classes can be considered~\cite{sm}. As shown below, this is indeed a good approximation. 

Let us define the probability to find the system in class ${\cal M}_0$, as $W_0(t) \equiv P_{k_0}(t)$. This is the survival probability of the initial state. The probability for being in the class ${\cal M}_1$ is $W_1(t) \equiv \sum_{k \in {\cal M}_1} P_{k }(t)$. Neglecting the back flow to the initial state,  we can write down the following set of rate equations~\cite{sm}, 
\begin{equation}
\begin{array}{lll}
\displaystyle \frac{dW_0}{dt} &= -\Gamma (W_0 - \overline{W_0^{\infty}}) , \\
&\\
\displaystyle \frac{dW_1}{dt} &= -\Gamma (W_1 - \overline{W_1^\infty}) +\Gamma(W_0- \overline{W_0^\infty}) ,\\
\label{casmod}
\end{array}
\end{equation}
where the infinite time averages are $\overline{W_0^\infty } = \sum_\alpha |C_{k_0}^\alpha|^4$ and
$\overline{W_1^\infty }= \sum_{k \in {\cal M}_1} \sum_\alpha |C_{k_0}^\alpha|^2 |C_{k}^\alpha|^2$. 
 
The decay rate $\Gamma$ 
corresponds to the width of the LDOS, 
\begin{equation}
F_{k_0} (E) =\sum_{\alpha}  | C_{k_0}^{\alpha}|^2 \delta (E - E^{\alpha }), 
\end{equation}
which is obtained by projecting the initial state $\ket{k_0}$ onto the energy eigenbasis. It  was introduced in nuclear physics  to describe the relaxation of  excited heavy nuclei~\cite{bohr}, where it is known as ``strength function''.

The solution of Eq.~(\ref{casmod}) gives 
\begin{equation}
\label{sur-cm}
\begin{array}{lll}
\displaystyle 
W_0(t) &= e^{-\Gamma t} (1- \overline{W_0^\infty}) + \overline{W_0^\infty} , \\
&\\
W_1(t) &= \Gamma t e^{-\Gamma t} (1-\overline{W_0^\infty} ) + \overline{W_1^\infty} (1-e^{-\Gamma t}). 
\end{array}
\end{equation}
With the expressions (\ref{sur-cm})   one can derive the time dependence for $N_{pc}(t)$, 
\begin{equation}
\label{pr-cm}
N_{pc}(t) \simeq \left[\sum_n  W_n^2/{\cal N}_n\right]^{-1} \! \simeq \left[ W_0^2 + W_1^2/{\cal N}_1 \! \right]^{-1} \sim e^{2\Gamma t},
\end{equation}
where ${\cal N}_n$ is the number of states contained in the $n$-th class.
This  result shows that the  number of basis states effectively participating in the evolution of the wave packet increases exponentially in time with the rate $2\Gamma$.   For a finite number of particles, this growth lasts until  the saturation given by Eq.~(\ref{t-ipr}).  We note that exponential instability was also studied in \cite{casati}, where the number of harmonics of the Wigner function was shown to increase exponentially fast in time.

\begin{figure}[htb]
\vspace{0.cm}
\includegraphics*[width=\columnwidth]{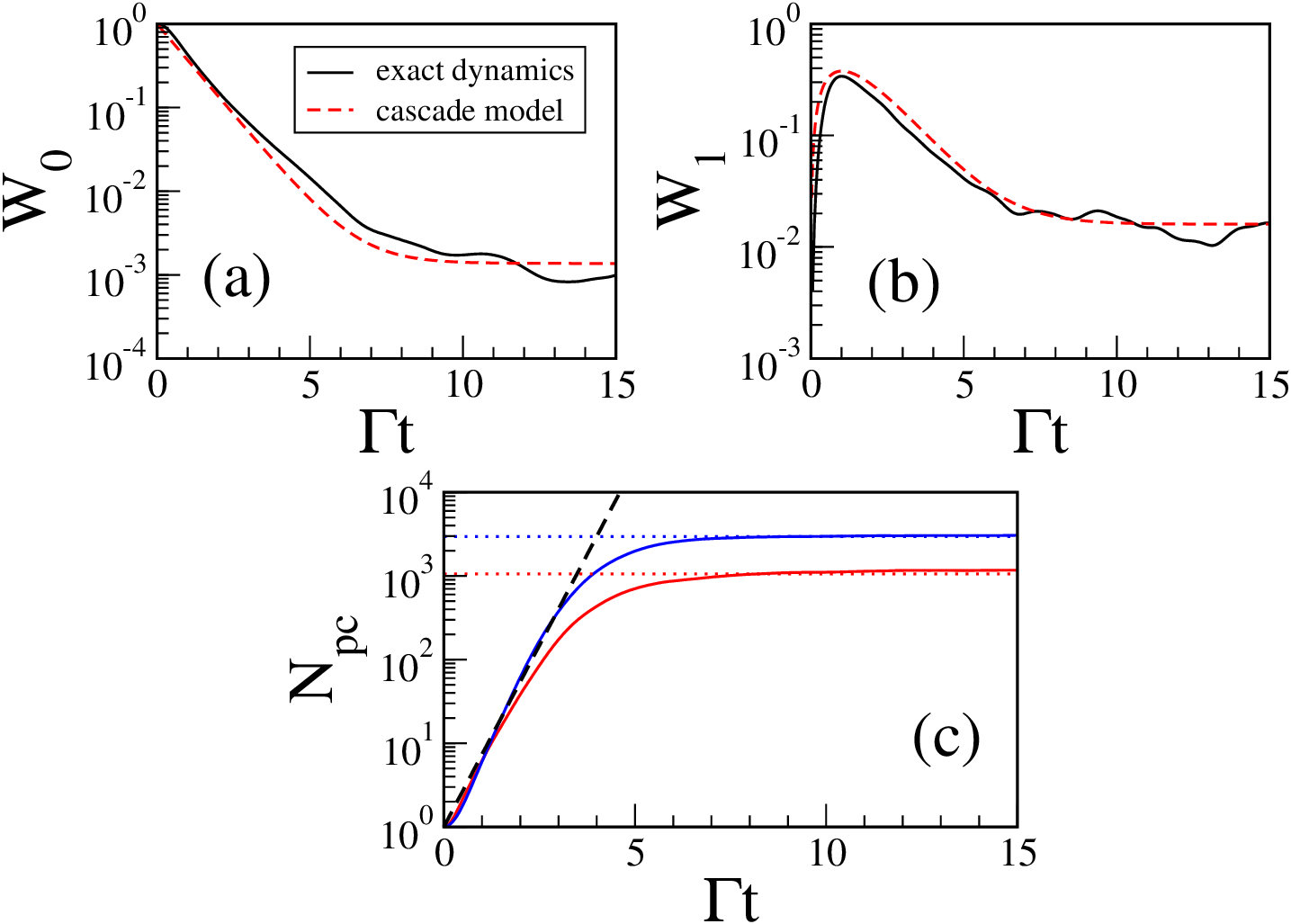}
\caption{ TBRI model: Numerical data  for $W_0(t)$ (a) and $W_1(t)$ (b) are shown by solid curves and compared with the analytical 
expressions~(\ref{sur-cm})  (dashed curves).   The parameters are $N=6$, $M=11$, $v=0.4$ (chaotic regime). In the initial state $\ket{\psi(0)} = (a^\dagger_5)^6\ket{0} $  all  particles initially occupy the 5-th single-particle level. The exponential rate $\Gamma = 2.8$ is obtained by fitting  $W_0(t)$. In (c): Growth in time of $N_{pc}$  for two initial conditions; from top to bottom:  $\ket{\psi(0)} = (a^\dagger_4 )^6\ket{0} $ and $\ket{\psi(0)} = (a^\dagger_5)^6\ket{0} $.  The dashed line is $e^{2\Gamma t}$.  Horizontal  dotted lines are the analytical estimates given by Eq.~(\ref{t-ipr}).  Average over $50$ random realizations.
}
\label{w0w1}
\end{figure}

{\em Results for the TBRI model.--}  To verify the validity of our approach, we compare in Fig.~\ref{w0w1}(a)
and (b)  the numerical data for $W_0(t)$ and $W_1(t)$ with Eqs.~(\ref{sur-cm}). The chosen $v$ is
such that the eigenstates are strongly chaotic and extended in the energy shell~\cite{bmi17}.
The value of $\Gamma$ used in the analytical expressions  is obtained by fitting the numerical curve for $W_0(t)$. 
The agreement between   numerical  and  analytical results is very good for the entire duration of the evolution, up to
the saturation given by  $\overline{W_0^\infty}$ and $\overline{W_1^\infty}$.  These results confirm that the back flow can indeed be neglected and that one can take into account two classes only.

In Fig.~\ref{w0w1}(c),  we show the evolution of the number of principal components $N_{pc}$. The numerical data (solid curve) corroborate the analytical prediction (dashed curve) from Eq.~(\ref{pr-cm}), namely
 the exponential behavior, $N_{pc}(t) \sim e^{ 2\Gamma t} $. 
 
Our data manifest the existence of two time scales. The first one, $t_\Gamma \simeq 1/\Gamma$, corresponds to  the characteristics decay time of $W_0(t)$, as shown in  Eq.~(\ref{sur-cm}). The second, $t_S$,
is the  time scale for the saturation of the dynamics and can be estimated 
from $e^{2\Gamma t} \simeq \overline{N_{pc}^\infty}$, which gives 
\begin{equation}
\label{ts}
t_S \simeq \ln (\overline{N_{pc}^\infty})/2\Gamma.
\end{equation}
Assuming a Gaussian shape for both the density of states and the LDOS~\cite{sm}, we show that  the maximal value of $\overline{N_{pc}^{\infty}}$ is
 \begin{equation}
\label{nmax}
 N_{pc}^{max}  = \eta \sqrt{1-\eta^2} {\cal D} 
\end{equation} 
where $\eta= \Gamma/\sigma \sqrt{2}$ and $\sigma $ is the width of the density of states (see details in SM~\cite{sm}).
For $M \sim 2N$ and for $M,N\gg 1$ one gets the estimate
\begin{equation}
\label{tsf}
t_S \sim  N/\Gamma = N t_\Gamma.
\end{equation}
This is the time scale for the complete thermalization in quantum MBS.
As one can see from Eq.~(\ref{tsf}), when the number of particles is very large, the two time scales are very different.
Notice that for fixed $\Gamma$, the time $t_S$ increases linearly with $N$ due to the exponential growth with $N$ of the many-body space and not because of the Gaussian shape of the density levels~\cite{sm}.
 
{\em Results for the spin model--} The analytical estimates obtained with the cascade approach are valid also for dynamical models. To show this, we study the evolution of the spin-1/2 system described by Eq.~(\ref{ham1}) in the limit of strong chaos ($\lambda =1$)~\cite{lea}. The analysis is analogous to the one developed with the TBRI model. We note, however, that $H_0$  is now initially written in the basis where each site has a spin pointing up or down in the $z$-direction (site-basis). 
It is then diagonalized to obtain the mean-field basis.  As a result, all matrix elements of the full Hamiltonian written in the mean-field basis become non-zero.
Therefore, to properly determine the classes,  we use the following procedure. In the first class we have 
all states $m$ coupled to $k_0$ such 
that $|H_{k_0, m}| > \xi |H_{k_0, k_0}-H_{m,m}|$  with $\xi$ being a threshold reasonably chosen. 
This procedure is repeated for higher classes.

\begin{figure}[htb]
\includegraphics*[width=\columnwidth]{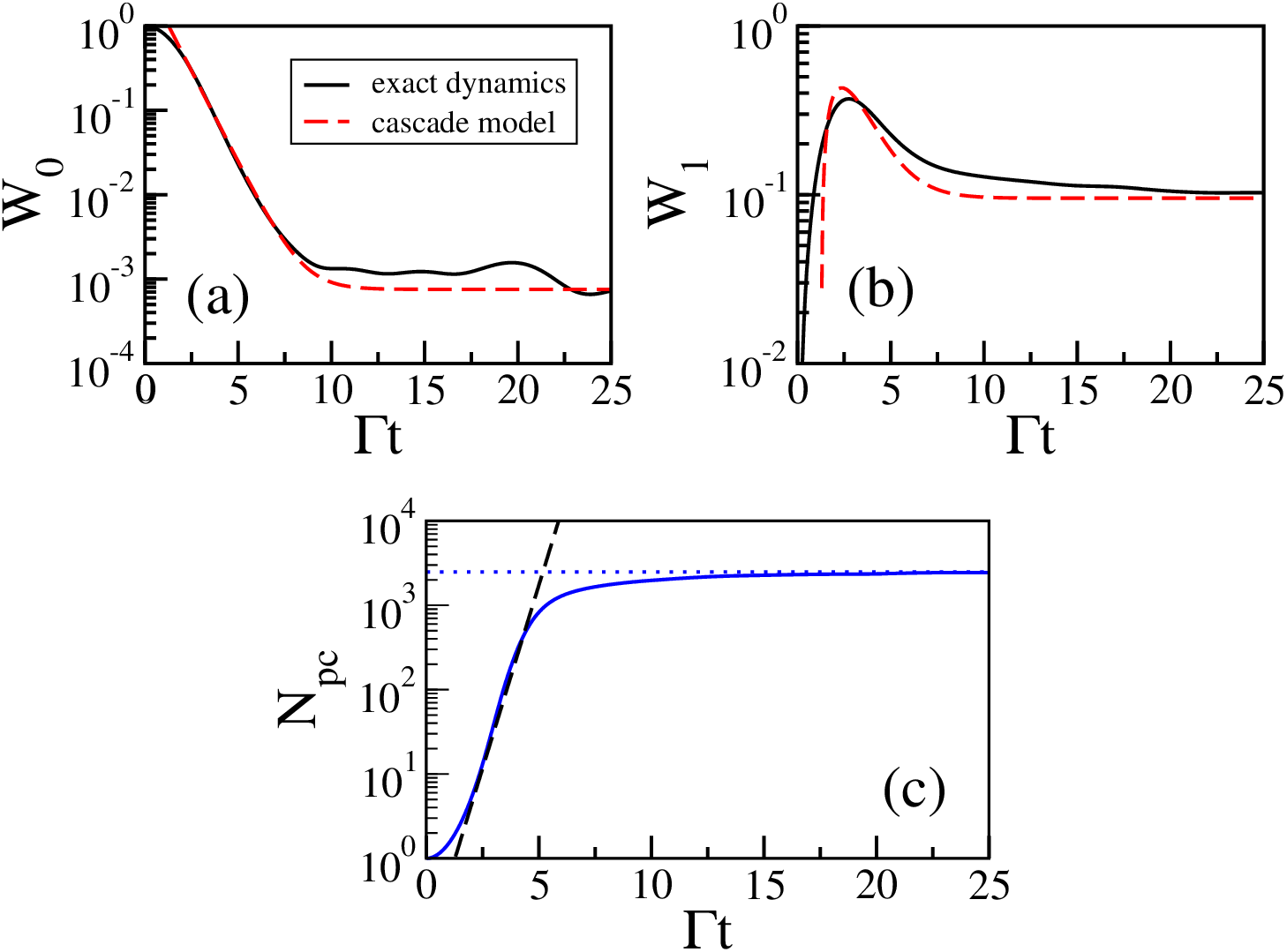}
\caption{Spin model: Numerical data (solid curves) for  $W_0(t)$ and   $W_1(t)$ 
compared with the analytical expressions~(\ref{sur-cm})  (dashed curves). 
In (c): Numerical data for the  number of principal components $N_{pc}(t)$ (solid curve) and  
  the infinite-time average in Eq.~(\ref{t-ipr}) (dotted line).  The dashed line
represents $e^{2\Gamma t}$.
 Parameters: $L=16$, $\Delta=0.48$, $\lambda=1$, and $N=7$ excitations. Average over 16 initial states with energy close to $-0.5$. Threshold for counting ${\cal N}_1$ is $\xi = 0.05$ and $\Gamma=2.62$ is obtained by fitting  $W_0 (t)$. }
\label{sm-all}
\end{figure}

Figure~\ref{sm-all} compares the numerical results for $W_0(t)$, $W_1(t)$, and $N_{pc}(t)$ for the spin model with the analytical expressions in Eq.~(\ref{sur-cm}) and Eq.~(\ref{pr-cm}). The agreement is very good and the exponential increase in time of the number of principal components with rate $2 \Gamma$ is  confirmed.
As for the TBRI model,  we  see   that the back flow is not important and that two classes suffice to describe the dynamics.
This validates our approach for realistic physical systems even in the absence of any random parameter.

{\em Discussion.--}
We studied the dynamics of interacting quantum MBS whose eigenstates have a chaotic structure in the basis of non-interacting particles.  We demonstrated that in the many-body space the relaxation is not a diffusive or ballistic process. Instead, wave packets evolve exponentially fast in the unperturbed basis before reaching saturation, which happens when all states of the energy shell get populated. Unexpectedly, we found that  the time scale for saturation is much  larger than the  characteristic decay time of the initial state. 

To describe the dynamical process, we developed a semi-analytical approach that allowed us to estimate the rate and the time scale of  the relaxation, as well as the saturation value of the number of principal components in the wave packet. It is quite impressive that our  simple phenomenological model with a single parameter -- the width $\Gamma$ of LDOS -- reproduces so well the system dynamics at very different time scales. 

The first analytical investigation of  the properties of  the LDOS was done by Wigner in his studies of banded random matrices~\cite{wigner}. In the context of quantum chaos, these matrices were employed in~\cite{chir}, where it was pointed out that the LDOS has a well defined classical limit and is the projection of the unperturbed Hamiltonian onto the total one. Its maximal width is given by the width of the energy shell, as shown in~\cite{chir}. In the classical description, the energy shell corresponds to the phase-space volume obtained by the projection of the phase-space surface $H_0 = E_0$ onto the surface defined by  the total Hamiltonian $H$. Note that the classical LDOS can be obtained by solving classical equations of motion~\cite{felix2000}. The dynamics of the classical packets created by $H_0$ is restricted to the energy shell~\cite{bgi98,felix2000}, which  can be filled in time either partially or ergodically.  In the quantum description, these two alternatives correspond to either localized or delocalized wave packets. 
  
Inspired by the above studies, our results for the exponential growth of $N_{pc}$  can be treated  in terms of the phase-space volume ${\cal V}_E$  occupied by the wave packet, ${\cal V}_E (t)\sim N_{pc }(t) /\rho(E) $, where  $\rho(E)$ is the total density of states. We can write 
\begin{equation}
\label{ks}
{\cal V}_E (t) = {\cal V}_E (0) e^{2\Gamma t} \sim {\cal V}_E (0)  e^{h_{KS}t}.
\end{equation}
Here, we associate $2\Gamma$ with the Kolmogorov-Sinai entropy~\cite{ksref}, $h_{KS}$, which gives the exponential growth rate of phase-space volumes for classically chaotic MBS~\cite{ksref}.  A connection between the entanglement entropy growth rate and $h_{KS}$ was found also in~\cite{bianchi}.
Note  that in many-body systems,  $h_{KS}$ is defined as the sum of {\it all } positive Lyapunov exponents and not only the largest one. The relation  $h_{KS} \sim 2\Gamma$ allows one to establish a quantum-classical correspondence for MBS. Indeed, when the system admits a well defined classical limit in which there is strong  chaos, 
the Kolmogorov-Sinai entropy is associated with the  width of the classical LDOS.

We stress that  Eq.~(\ref{ks})  holds only up to the saturation time $t_S \sim  N t_\Gamma$, which defines the time scale for the quantum-classical correspondence for the number of principal components $N_{pc}$ participating in the dynamics. This time $t_S$ is important for the problem of thermalization in isolated systems of interacting particles. It establishes the time scale for the complete thermalization of the system due to the ergodic filling of the energy shell. It also corresponds to the scrambling time discussed in studies of the loss of information in black holes (see \cite{scramble} and references therein). One sees that in the thermodynamic limit, $N \to \infty$, $t_S$ diverges (provided the width of the LDOS remains constant), which agrees with the quantum-classical correspondence principle.

{\em Acknowledgements.}--
We acknowledge discussions with G.~L. Celardo. F.B.  acknowledges support by the I.S. INFN-DynSysMath. FMI acknowledge
financial support from VIEP-BUAP Grant No. IZF-EXC16-G.
LFS was funded by the American National Science Foundation (NSF) Grant No. DMR-1603418.



\vspace{0.5cm}
\onecolumngrid

\begin{center}
{\large \bf Supplementary material for EPAPS\\Exponentially fast dynamics of chaotic many-body systems}

\vspace{0.6cm}
 
Fausto Borgonovi$^1,2$, Felix M. Izrailev$^3,4$, and Lea F. Santos$^5$

$^1${\it Dipartimento di Matematica e
  Fisica and Interdisciplinary Laboratories for Advanced Materials Physics,
  Universit\`a Cattolica, via Musei 41, 25121 Brescia, Italy}
  
$^2${\it Istituto Nazionale di Fisica Nucleare,  Sezione di Pavia,
  via Bassi 6, I-27100,  Pavia, Italy}

$^3${\it Instituto de F\'{i}sica, Benem\'{e}rita Universidad Aut\'{o}noma
  de Puebla, Apartado Postal J-48, Puebla 72570, Mexico}
  
$^4${\it Dept. of Physics and Astronomy, Michigan State University, E. Lansing, Michigan 48824-1321, USA}

$^5${\it Department of Physics, Yeshiva University, New York,
New York 10016, USA}

\end{center}

\section{ Dynamics in the many-body space} 

\begin{figure}[ht!]
\vspace{0.cm}
\includegraphics*[width=12cm]{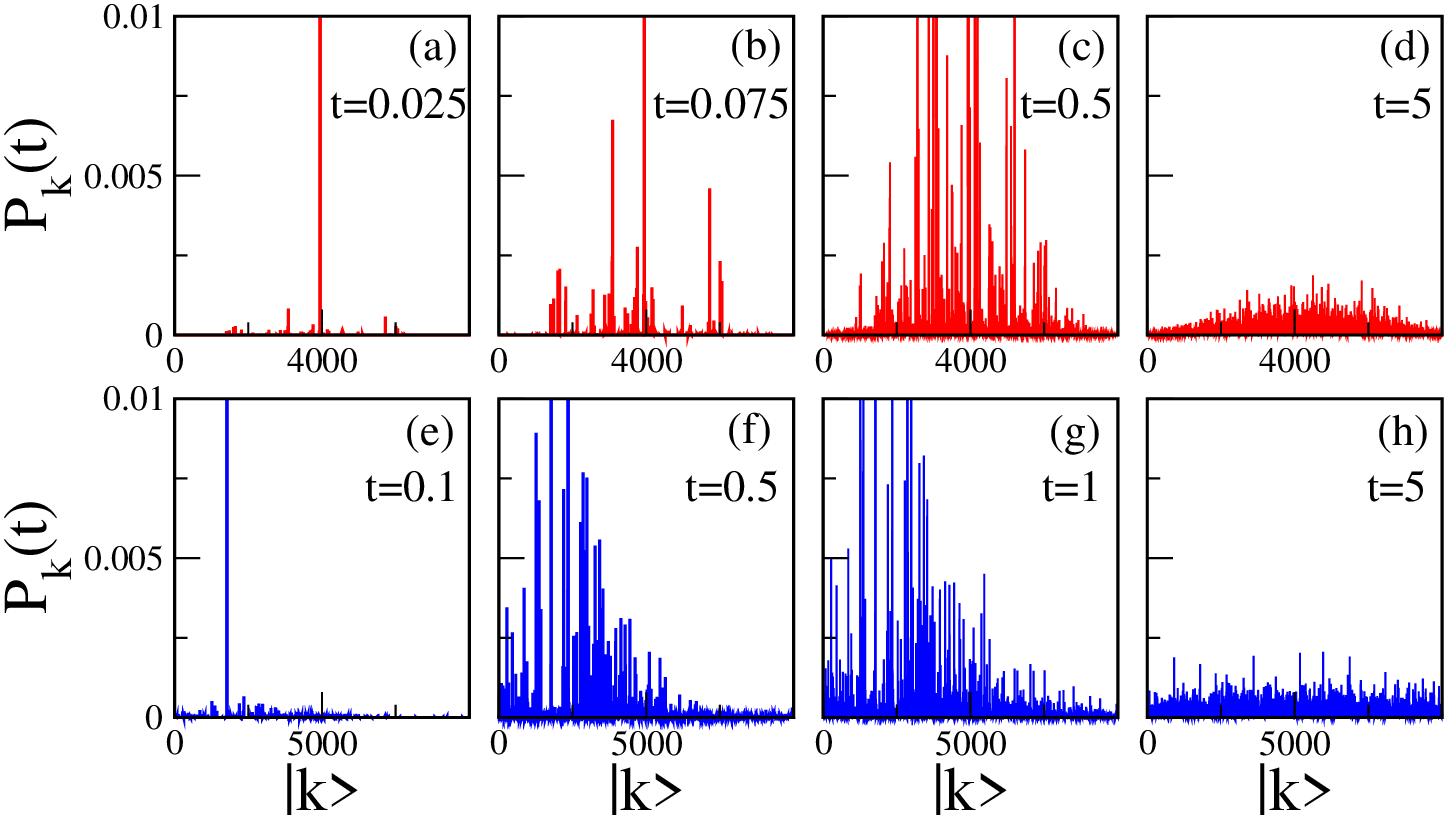}
\caption{Spread of the probability $P_k(t)$ onto the unperturbed basis states $\ket{k}$ for fixed times, indicated in the panels, for the TBRI model [(a)-(d)] and for the spin model [(e)-(h)]. For the TBRI model, $N=6$, $M=11$, $v=0.4$  and one realization of random potential. For the spin model, $L=16$, $7$ excitations, $\Delta=0.48$, $\lambda=1$, open boundaries. The energy of the initial basis state is chosen close to the middle of the spectrum.}
\label{wf}
\end{figure}

 A main problem when studying the dynamics of systems with many interacting particles is that it cannot be described as either ballistic or diffusive in the many-body space. Instead, it is the initial unperturbed many-body state that spreads onto other unperturbed many-body states in a complicated way. A pictorial demonstration of how this happens is given in Fig.~\ref{wf} ,  where we show $P_k(t) = |\langle k|\psi(t) \rangle|^2$ as a function of the unperturbed state $|k\rangle$ for different times for both the TBRI model (top panels (a)-(d)) and the  the dynamical spin model (bottom panels (e)-(h)).  The process is equivalent to the flow of probabilities through different sets of the unperturbed many-body states.
In a small time scale, only the basis vectors directly coupled to the initial state are excited [Fig.~\ref{wf} (a) and (e)]. The number of these states  is much smaller than the total number of basis vectors, which is a consequence of the sparse character of the Hamiltonian matrix. For longer times, as shown in Fig.~\ref{wf} (b), (f) and Fig.~\ref{wf} (c), (g), the states participating in the dynamics sparsely fill a large portion of the many-body space that is within the energy shell. As time passes, more basis states are populated inside the shell, until it gets ergodically filled  [Fig.~\ref{wf} (d), (h)]. This ergodic filling takes place provided the perturbation $V$ is sufficiently strong, so that the eigenstates of $H$ are delocalized in the energy shell.

\section{Infinite time average of the number of principal components}

The purpose of this section is to find an estimate of 
\begin{equation}
\label{t-ipr}
[ \, \overline{N_{pc}^{\infty}} \,]^{-1}  =    2 \sum_k (P_{k}^d)^2  -  \sum_{\alpha}  | C_{k_0}^\alpha |^4 \sum_k |C_{k}^\alpha|^4  
\end{equation}
in terms of fundamental characteristics  of the system, {\em i.e.}   without the explicit diagonalization of the full Hamiltonian matrix.

\begin{figure}[ht!]
\vspace{0.cm}
\includegraphics*[width=12cm]{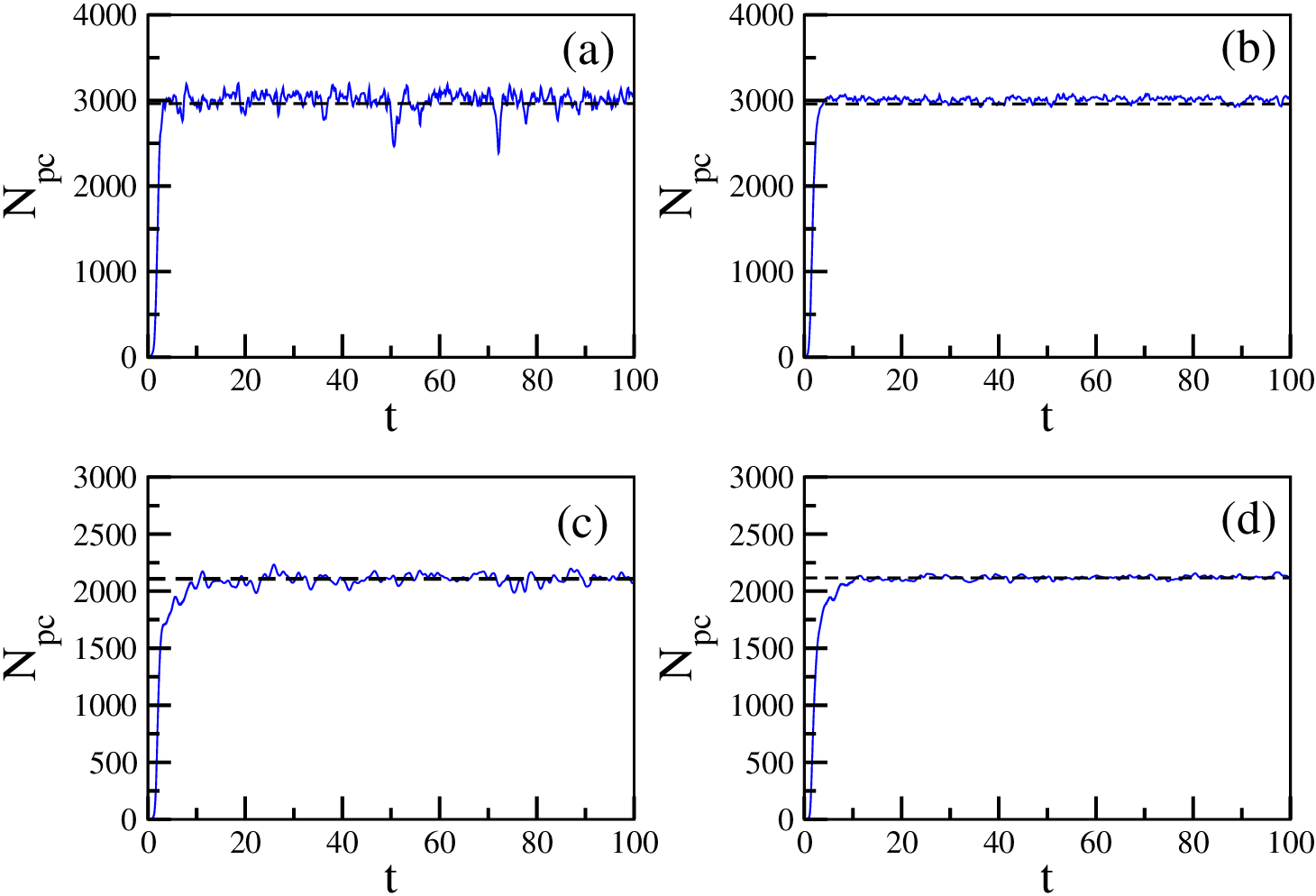}
\caption{Number of principal components $N_{pc}$ as a function of time in normal scale (corresponding to the blue curve of Fig.1 (c) and of Fig.2 (c) of the main text). Top panels: TBRI model with only one realization of the random potential (a) and with 10 random realizations (b); $N=6$, $M=11$, $v=0.4$.
Bottom panels: spin model for a single initial state (c) and averaged over five initial states (d); $L=16$, $7$ excitations, $\Delta=0.48$, $\lambda=1$, and open boundaries.} 
\label{flu}
\end{figure}

To start with, we notice that the second term in the r.h.s of Eq.~(\ref{t-ipr}) is roughly $1/{\cal D}$ times smaller than the first one. This can be seen by taking uncorrelated components
$ C_{k}^\alpha  \simeq ({1/\sqrt{\cal D}}) e^{i\xi_{\alpha,k}}$, where $ \xi_{\alpha,k}$ are random numbers. Thus 
\begin{equation}
\label{t-ipr1}
 2 \sum_k (P_{k}^d)^2 = 2\sum_{\alpha,\beta, k}  | C_{k_0}^\alpha |^2  |C_{k}^\alpha|^2 | C_{k_0}^\beta |^2  |C_{k}^\beta|^2 ,
\simeq \frac{{\cal D}^3}{{\cal D}^4} \simeq \frac{1}{\cal D}
\end{equation}
while
\begin{equation}
\label{t-ipr2}
\sum_{\alpha,k }  | C_{k_0}^\alpha |^4  |C_{k}^\alpha|^4 \simeq \frac{{{\cal D}^2}}{{\cal D}^4} \simeq \frac{1}{{\cal D}^2}
\end{equation}
We can then take the first term only,
\begin{equation}
\label{t-ipra}
\left[\, \overline{N_{pc}^{\infty}}\, \right]^{-1} \simeq  2 \sum_k (P_{k}^d)^2 .
\end{equation}
Of course, in the real dynamics of finite systems, the fluctuations around the long-time average will always be present.
To reduce them, we routinely average over random configurations for the TBRI and over initial states close in energy for the spin model. In Fig.~\ref{flu}, we compare the sizes of the temporal fluctuations for a single realization (a) and a single initial state (c) with the averages in (b) and (d).

\begin{figure}[ht!]
\vspace{0.cm}
\includegraphics*[width=12cm]{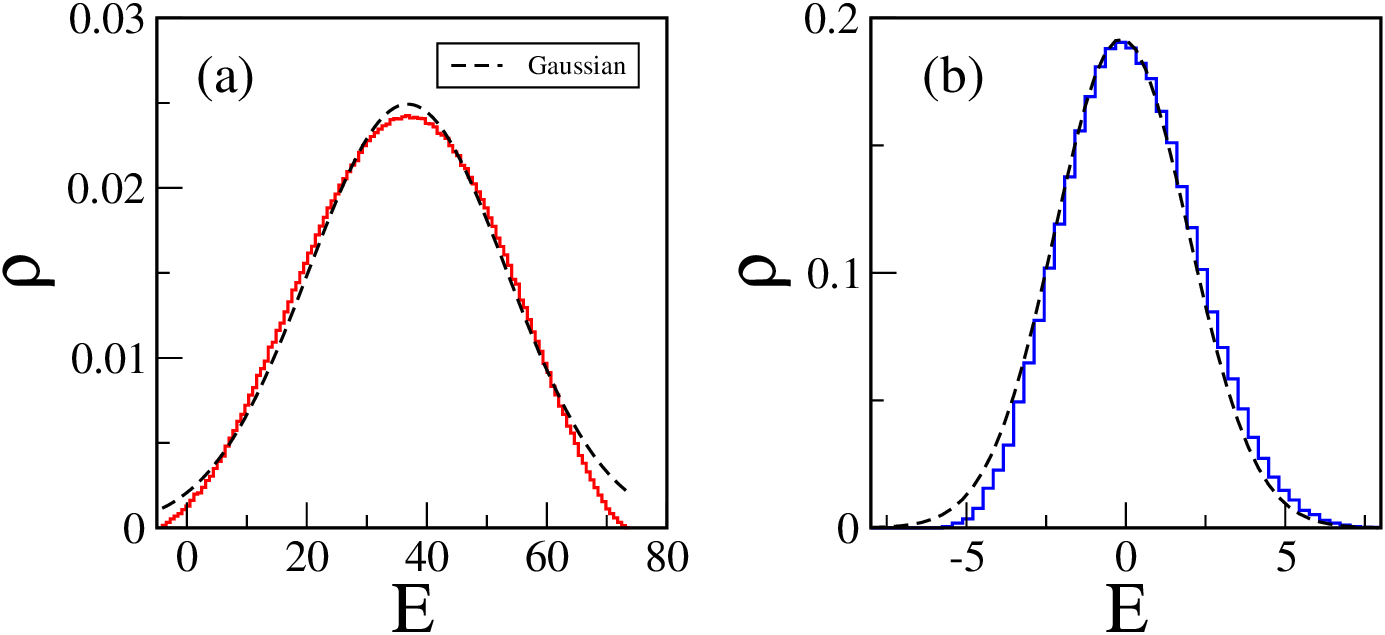}
\caption{Density of states for the TBRI model for 10 realizations (a) and for the spin model with a single realization (b). The dashed line is a Gaussian. Here, $N=6$, $M=11$, $v=0.4$  (a) and  $L=16$, $7$ excitations, $\Delta=0.48$, $\lambda=1$, 
and open boundaries (b). }
\label{f-dos}
\end{figure}

Let us now assume a Gaussian shape for (i) the LDOS, (ii) the density of the unperturbed states, and (iii) the density of states of the full Hamiltonian. This is a realistic assumption for chaotic many-body systems with two-body interactions, such as the TBRI model (see Ref.~[23] of the main text) and the spin model (see Ref.~[17] of the main text). In Fig.~\ref{f-dos} we show the density of states for both models. As one can see, the Gaussian approximation is very good.

(i) For the LDOS we then have 
\begin{equation}
\label{ldos}
F_k(E) = \sum_{\alpha }  | C_{k}^\alpha |^2  \delta(E-E^\alpha)  \simeq  \frac{1}{\Gamma\sqrt{2\pi}} 
\exp \left\{ -\frac{(E-E_k^0)^2}{2\Gamma^2} \right\} ,
\end{equation}
where $\Gamma$ is the width of the LDOS and $E_k^0 $ is the energy of the unperturbed state. We assume that $\Gamma$ is independent of $E_k^0 $.
The LDOS is normalized, $\int dE F_k(E) = 1$. 

(ii) The Gaussian shape  for  the unperturbed density of states $\rho_0(E)$ of width $\sigma_0$ is written as
\begin{equation}
\label{dos0}
\rho_0(E) =     \frac{{\cal D}}{\sigma_0\sqrt{2\pi}} 
\exp \left\{ -\frac{E^2}{2\sigma_0^2} \right\} .
\end{equation} 

(iii) 
The Gaussian density of states, characterized by a width $\sigma$, is such that 
\begin{equation}
\label{dos}
\rho(E) =     \frac{{\cal D}}{\sigma\sqrt{2\pi}} 
\exp \left\{ -\frac{E^2}{2\sigma^2} \right\} ,
\end{equation}
where for simplicity we set the middle of the spectrum at the energy $E=0$.
The density of states is normalized to the dimension of the many-body space, $\int dE \rho(E) = {\cal D}$.

The above assumptions imply that in the continuum, one has 
\begin{equation}
\label{pkd}
P_{k}^d = \sum_{\alpha}  | C_{k_0}^\alpha |^2  |C_{k}^\alpha|^2  \simeq  \int \, dE \, \rho(E)^{-1} F_k(E) F_{k_0} (E)
\equiv {\cal G}_{k_0} (E_k^0),
\end{equation} 
  where the function
\begin{equation}
\label{gkk0}
{\cal G}_{k_0} (E_k^0) = \frac{\sigma^2}{\Gamma{\cal D}\sqrt{2\sigma^2-\Gamma^2}} 
\exp \left\{ 
-\frac {(E_k^0)^2+(E_{k_0}^0)^2} {2\Gamma^2}  +   
\frac {(E_k^0+E_{k_0}^0)^2} {2\Gamma^2(2\sigma^2-\Gamma^2)}   
\right\}
\end{equation} 
is defined only for $2\sigma^2>\Gamma^2$. We can then approximate
\begin{equation}
\label{t-ipra1}
[ \, \overline{N_{pc}^{\infty}} \,] ^{-1} \simeq  2 \sum_k (P_{k}^d)^2  \simeq  2\int \, dE  \, \rho_0(E) {\cal G}_{k_0} (E)^2.
\end{equation}
Taking into account that $\sigma_0^2 = \sigma^2 - \Gamma^2$ (see Ref.~[23] of the main text), Eq.~(\ref{t-ipra1}) gives
\begin{equation}
\label{prf}
 \overline{N_{pc}^{\infty}}  =    {\cal D} \frac{\Gamma\sqrt{2\sigma^2-\Gamma^2}}{2\sigma^2} e^{-E_{k_0}^2/\Gamma^2} .
\end{equation}

The maximal value of $\overline{N_{pc}^{\infty}}$ occurs in the middle of the energy spectrum, where $E_{k}^0 = 0$. Defining $\eta = \dfrac{\Gamma}{\sqrt{2}\sigma}$, we have
\begin{equation}
\label{eta} 
 \overline{N_{pc}^{max} }= \frac{\Gamma}{\sqrt{2}\sigma} \sqrt{ 1 - \left( \frac{\Gamma}{\sqrt{2}\sigma}   \right)^2} {\cal D} = \eta \sqrt{1 - \eta^2} {\cal D} \equiv 
\Xi (\eta) {\cal D} ,
\end{equation}
One sees that Max$_{\eta} \left[ \Xi (\eta) \right] = 1/2$, so
\begin{equation}
\label{etamax}
 \overline{ N_{pc}^{max} } \leq   {\cal D}/2.
\end{equation}

\section{Cascade Model}
 
The dynamics of a chaotic quantum many-body system with two-body interaction, where many chaotic eigenstates are present, can be approximated in the following way. We introduce subclasses for all basis states in such way that: (i) the ${\cal M}_0$ class contains only the initial state $\ket{k_0}$, (ii) the class ${\cal M}_1$ contains only the ${\cal N}_1$ states that are directly coupled with initial basis state , i.e. all those $|k\rangle$ for which $\langle k_0 | V | k\rangle \ne 0$, (iii) the second class  ${\cal M}_2$  contains all $\ket{f}$ states such that $\langle k_0 | V | j\rangle \langle j | V | f\rangle \ne 0$  for some $j$, and so on.

It is clear that initially, at $t=0$, only the class  ${\cal M}_0$ is populated. As the time passes, the first class starts being populated, then the second class, and so on, as illustrated in Fig.~\ref{wf}. The idea is now to write down a set of probability conservation equations to describe this picture, such as  
\begin{equation}
\begin{array}{lll}
\displaystyle \frac{dW_0}{dt} &= -\Gamma (W_0 - \overline{W_0}) , \\
&\\
\displaystyle \frac{dW_1}{dt} &= -\Gamma (W_1 - \overline{W_1}) +\Gamma(W_0- \overline{W_0}) ,\\
&\\
\displaystyle \frac{dW_2}{dt} &= -\Gamma (W_2 - \overline{W_2}) +\Gamma(W_1- \overline{W_1}) ,\\
&\\
...
\label{casmod}
\end{array}
\end{equation}
Above, $W_0(t)$ is simply the survival probability, that is the probability of being in the initial state. The survival probability is the square of the Fourier transform of the LDOS (see {\em e.g.} Ref.[17] of the main text). Therefore, the typical decay time of the survival probability is the inverse of the width $\Gamma$ of the LDOS, which explains the first line in Eqs.~(\ref{casmod}). There are two necessary elements for achieving the whole set of equations (\ref{casmod}). The first is that a single parameter ($\Gamma$) suffices to describe the entire probability flow from one class to the other. This holds when the eigenstates are chaotic. The second is that the values $\overline{W_k}$ need to added, because in any finite system, the dynamics saturates to a finite value, so the long-time probability to be in some class cannot be zero. 

In writing Eq.~(\ref{casmod}), we also assume that the probability of the return from class ${\cal M}_{k+1} $ to the previous class ${\cal M}_{k} $ is small. This is a valid approximation, since the number of states ${\cal N}_{k+1}$ is much larger than the number of states ${\cal N}_{k}$. A rough estimate gives ${\cal N}_{k} \approx M^k$.  Note that we consider a system that is far from equilibrium. Evidently, if the system was at equilibrium, the probabilities for all states within the energy shell  would be of the same order.

In any finite system, the number of classes is finite, so the set of equations can be closed by setting the conservation of probability to the last class considered, such that
$$
W_f = 1 -\sum_{k=0}^{f-1} W_0.
$$
To be consistent with the cascade model, $f$ needs to be chosen in such a way that ${\cal N}_{f-1} \ll {\cal D}$,
where ${\cal D}$ is the dimension of the many-body (Hilbert) space. For the values of $N$ and $M$ that we use,  the number of states in the second class already coincides with ${\cal D}$ . For this reason, we restrict our analysis to two classes only.  This is not a restriction of our model, but simply of our computers. To have more than two classes in fully chaotic systems, we need $N \gg M \gg 1 $, which is beyond our computer capabilities.

We stress that the validity of Eqs.~(\ref{casmod}) has been checked numerically against the real Schr\"odinger dynamics for both the TBRI model and the spin model in Figs. 1 (a), (b) and Figs. and 2 (a), (b) of the main text. Our numerical results legitimate our equations. Beyond the perturbative regime, they describe accurately the dynamics all the way to saturation.

\section{Time scales}

As it is clear from the solution of the cascade model for the survival probability,
$$
W_0(t) = e^{-\Gamma t} (1- \overline{W_0^\infty}) + \overline{W_0^\infty} .
$$
The width $\Gamma$ is related to the time scale,
$$
t_\Gamma \simeq \frac{1}{\Gamma}
$$
for the depletion of $W_0$. This means that after the time $t_\Gamma$, the probability to be in the class ${\cal M}_0$ is reduced by the factor $1/e$.
 
We have also seen that global observables, such as the number of principal components $N_{pc}$, grow
exponentially in time,
$$
N_{pc} (t) \simeq e^{2\Gamma t} ,
$$
up to the saturation point given by $ \overline{N_{pc}^{\infty}} $. It is quite natural to estimate the saturation time $t_S$ as the time for which
$$
e^{2\Gamma t_S} \simeq \overline{N_{pc}^{\infty}} ,
$$
Using Eq.~(\ref{eta}),
\begin{equation}
\label{th}
t_S \simeq \frac{1}{2\Gamma} \ln \left[\Xi ({\eta}) {\cal D} \right].
\end{equation}
To evaluate the dimension of the Hilbert space, we should distinguish between the TBRI and the spin model.

$\bullet $ The dimension of the Hilbert space for the TBRI model can be expressed in terms of the number of bosons $N$ and
the number of single particle energies $M$, as 
$$
{\cal D} = \frac{(N+M-1)!}{N!(M-1)!}.
$$
In the  limit of $N, M \gg 1$, using the Stirling approximation,
one has 
\begin{equation}
\ln {\cal D} \approx  N\ln \left( 1+\frac{M}{N}\right) +  M\ln \left( 1+\frac{N}{M}\right).
\label{sti}
\end{equation}
In the dilute limit, $M\simeq 2N$, and for $N,M \gg 1$, 
one finally gets the estimate for the saturation time,
\begin{equation}
\label{thf}
t_S = \frac{\ln\Xi ({\eta}) }{2\Gamma}  +c_1\frac{N}{\Gamma} ,
\end{equation}
where $c_1 = \ln(27/4)$ is a constant of order 1.
Since the first term on the r.h.s term above is independent of the number of particles $N$, we have
that in the thermodynamic limit and for a fixed ratio $N/M$ ,
$$
t_S \approx  N t_\Gamma .
$$

$\bullet $ For the spin model with $N$ excitations in $L$ different sites, one has 
$$
{\cal D} = \frac{L!}{N!(L-N)!}.
$$
In the  limit of $N, L \gg 1$, using the Stirling approximation,
one gets
\begin{equation}
\ln {\cal D} \approx  N\ln \left( \frac{L}{N}-1\right) - L\ln \left( 1-\frac{N}{L}\right).
\label{sti1}
\end{equation}
For half-filling, $L \simeq 2N$, and for $N,L \gg 1$, 
one finally obtains the estimate for the saturation time,
\begin{equation}
\label{thf1}
t_S = \frac{\ln\Xi ({\eta}) }{2\Gamma}  +c_2\frac{N}{\Gamma} ,
\end{equation}
where $c_2 = \ln(4)$ is a constant of order 1.
Since the first term on the r.h.s term above is independent of the number of excitations $N$, we have
that in the thermodynamic limit and for a fixed ratio $N/L$ ,
$$
t_S \approx  N t_\Gamma .
$$

It is important to stress that this estimate does not depend on the exact shape of the LDOS and density of states. What is essential here is that the size of the Hilbert space ${\cal D}$ increases exponentially with the number of particles $N$ and that $\overline{N_{pc}^\infty} \propto {\cal D} $, which is indeed the case for typical quantum many-body systems.

We notice the similarity between the results for the many-body case and for the one-body quantum chaotic system (the kicked rotor model, discussed in the introduction of the main text). Both models exhibit two different time scales. Here we have  $t_S \gg t_\Gamma$. For the one-body case, we have the Ehrenfest time $t_E$ and a much longer diffusive time $t_D$. $t_E$ is related to the initial exponential spreading of the wave packets and is linked with the divergence of neighboring trajectories in the phase space of classically chaotic systems. Instead, $t_D$ is associated with the quantum energy diffusion, in analogy with the classical unbounded diffusion.

\end{document}